# Measurement of the absolute wavefront curvature radius in a heterodyne interferometer

**Gerald Hechenblaikner[1,*]**

[1] *Gerald Hechenblaikner, EADS Astrium, Friedrichshafen, Germany*

[*]*Corresponding author: Gerald.Hechenblaikner@astrium.eads.net*

We present an analytical derivation of the coupling parameter relating the angle between two interfering beams in a heterodyne interferometer to the differential phase-signals detected by a quadrant photo-diode. This technique, also referred to as Differential Wavefront Sensing (DWS), is commonly used in space-based gravitational wave detectors to determine the attitude of a test-mass in one of the interferometer arms from the quadrant diode signals. Successive approximations to the analytical expression are made to simplify the investigation of parameter dependencies. Motivated by our findings, we propose a new measurement method to accurately determine the absolute wave-front curvature of a single measurement beam. We also investigate the change in coupling parameter when the interferometer "test-mirror" is moved from its nominal position, an effect which mediates the coupling of mirror displacement noise into differential phase-measurements.

# Introduction

In applications of optical metrology and instruments it is important to characterize not only the intensity but also the phase-distribution of the beams propagating through the system. Wavefront sensing [1], that is the measurement and characterization of the opical wavefront, also lies at the heart of Adaptive Optics [2], where one tries to compensate a wavefront aberration or mismatch of wavefront curvature [3] through iterative adjustment of corrective optical elements. The wavefront may also be used as a probe of the surface roughness and curvature of an optical element from which the probe beam is reflected. Common tools for wavefront sensing are the Hartmann-Shack sensor, which allows reconstruction of the complete waveform from gradient and Laplacian data [4], or phase-shifting-interferometry [5].

In this paper we propose a novel method of determining the absolute wavefront curvature of one of the interfering beams in a heterodyne interferometer. Optical heterodyne interferometers find numerous applications in the field of high-precision metrology where distance variations are measured with sub-wavelength precision over a large dynamic range [6, 7]. The heterodyne interferometers used in gravitational wave detection technology under current development [8-13] are used to not only precisely determine the position of a test-mass but also its attitude, to pico-meter and nano-rad precision, respectively. The test-mass attitude is determined from differential wavefront sensing (DWS) signals. In differential wave-front sensing the relative angle between two interfering beams is inferred from the differential phase across the beam. The latter is usually obtained from the signals detected by the two halves of a photo-diode. The great sensitivity of differential wave-front sensing coupled with its simple and practical implementation has made it the method of choice for attitude measurements in gravitational wave-detectors.

This paper is structured as follows: We first derive an analytical expression for the coupling factor relating the attitude of a test-mirror to the detected DWS-signals. Starting from this expression we make a series of successive approximations which allow us to gain more insight into the beam parameter dependencies of the coupling factor. Based on our findings we propose a novel method of directly measuring the absolute wavefront curvature of the measurement beam; further image processing or data analysis is not required. We then investigate the dependency of the coupling parameter on translational movements along the optical path axis of the test-mirror. These results are used to discuss the coupling of longitudinal mirror-jitter into angular phase measurements in a parameter regime typical for space-borne heterodyne interferometers.

**Experimental schematic**

A simplified schematic of a heterodyne interferometer is depicted in Figure 1. The output of a laser is split into two beams, termed "reference" and "measurement" beam. The two beams are frequency-shifted relative to another by the heterodyne frequency $f_{het}$ in the following accousto-optic-modulators AOM1 and and AOM2. To clean up the beam profile the beams are then fed through optical fibres and fibre injectors onto the optical bench. There the measurement beam is following a path where it is reflected from two rotatable mirrors before being recombined with the reference beam on the recombination beam splitter. The interference pattern on the photo-diode is beating with the heterodyne frequency $f_{het}$ and its phase is detected and measured by the two halves of a quadrant photo-diode and a phase-meter.

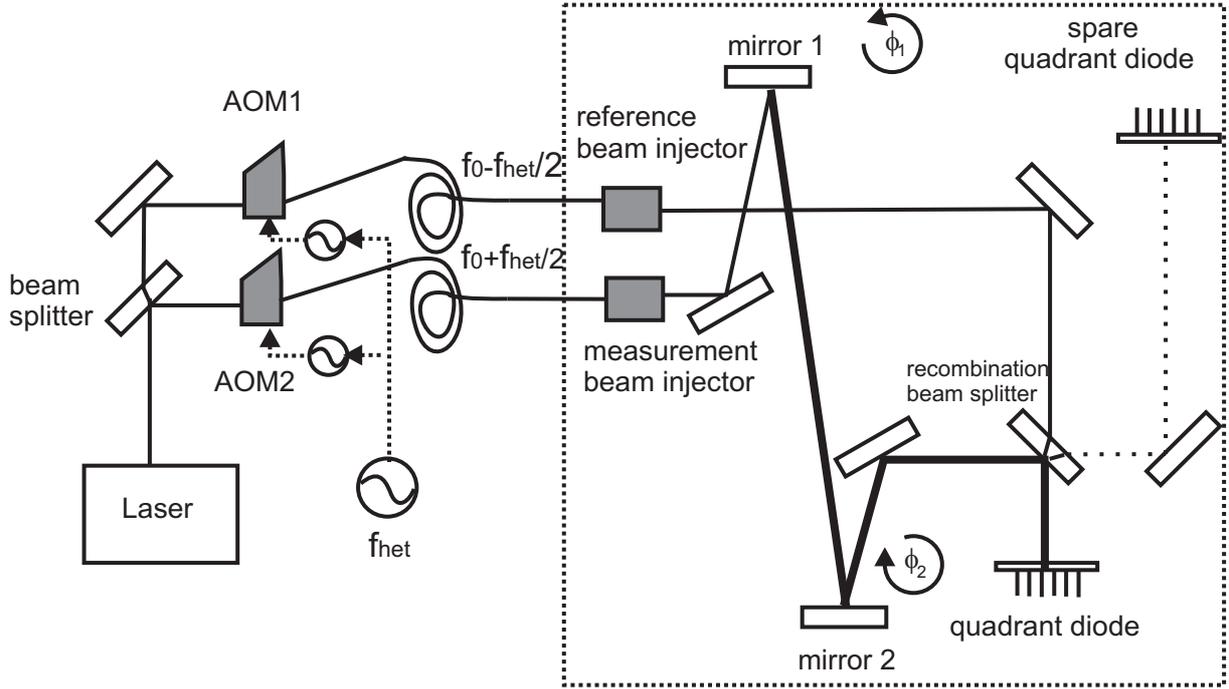

**Figure 1: The experimental setup to measure the absolute curvature of the measurement beam wave-front. Mirror 1 and mirror 2 are successively rotated in the opposite sense of rotation. At thick line is printed to indicate the lever arm from mirror 1 to the photo-diode. The path to the spare photo-diode is printed as a dotted line.**

## Analytical Derivation of the DWS coupling parameter

In this section we derive an analytic expression for the differential phase of the interference pattern detected by the quadrants of the photo-diode. The amplitude of a Gaussian beam propagating along the z-direction beam is given by the following expression

$$A = \frac{1}{w(z)}\sqrt{\frac{2}{\pi}} e^{-(x^2+y^2)/w^2(z)} e^{ik(x^2+y^2)/(2R)} e^{ikz} \qquad (1)$$

where the beam waist $w(z)$ and the beam radius of curvature $R(z)$ at position $z$ are given in terms of the beam waist $w_0$ and the Rayleigh range $z_R = \pi w_0^2 / \lambda$ by

$$w^2(z) = w_0^2 \left(1 + \frac{z^2}{z_R^2}\right)$$
$$R(z) = z\left(1 + \frac{z_R^2}{z^2}\right) \tag{2}$$

We consider the case where two beams, the "measurement" and the "reference beam", are interfering on a quadrant photo-diode, as commonly used in interferometry, and calculate the differential phase in the horizontal direction (analogue for the vertical direction).
The two beam amplitudes are given in their respective reference frames by:

$$A_r = C_r e^{-\frac{x^2 + y^2}{w_r^2(z)}} e^{ik\frac{x^2 + y^2}{2R_r(z)}} e^{ikz}$$
$$A_m = C_m e^{-\frac{\tilde{x}^2 + \tilde{y}^2}{w_m^2(\tilde{z})}} e^{ik\frac{\tilde{x}^2 + \tilde{y}^2}{2R_m(\tilde{z})}} e^{ik\tilde{z}}, \tag{3}$$

where the constants $C = w^{-1}(z) \cdot (2/\pi)^{1/2}$ have been introduced for convenience. They intentionally are omitted from this point on as they have no impact on phase-differences and cancel out. The two beams propagate along the $z$-axis of their respective reference frames. For simplicity and without loss of generality we assume that the reference beam is centered in the frame of the quadrant diode ($x = 0$), but the center of the measurement beam is offset by a distance $x_0$. The profiles of the two beams on the interference plane and their respective reference frames are depicted in Figure 2.

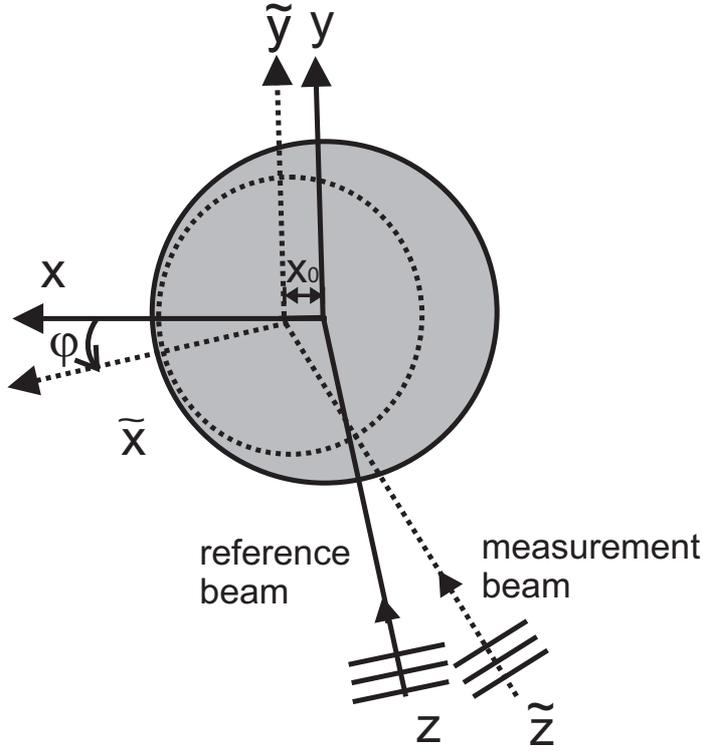

**Figure 2: The reference frames of the measurement beam (axes with tilde) and the reference beam (axes without tilde). The two beam propagate along $\tilde{z}$ and $z$, respectively, and interfere on the plane described by $z=0$.**

The reference frame of the measurement beam $(\tilde{x}, \tilde{y}, \tilde{z})$ is obtained from the reference frame of the reference beam through a simple rotation and a translation along the x-axis:

$$\begin{pmatrix} \tilde{x} \\ \tilde{y} \\ \tilde{z} \end{pmatrix} = \begin{pmatrix} \cos\varphi & 0 & -\sin\varphi \\ 0 & 1 & 0 \\ \sin\varphi & 0 & \cos\varphi \end{pmatrix} \begin{pmatrix} x \\ y \\ z \end{pmatrix} - \begin{pmatrix} x_0 \\ 0 \\ 0 \end{pmatrix} \qquad (4)$$

Assuming that the angle $\varphi$ between the two beams is small ($<1 mrad$), we neglect terms of order $\varphi^2$ or higher and obtain an expression for the amplitude of the measurement beam in the coordinate frame of the reference beam:

$$A_m = e^{-\frac{(x-x_o)^2+y^2}{w_m^2(z)}} e^{ik\frac{(x-x_o)^2+y^2}{2R_m(z)}} e^{ikz} e^{ikx\cdot\varphi} \quad (5)$$

Note that the photo-diode quadrants act as integrators which average the amplitudes of the incident beams. We now calculate the average complex amplitude of the interference pattern on the left half of the quadrant diode which is centered at $z = 0$:

$$F_{left} = \int_{-\infty}^{\infty} dy \int_{-\infty}^{0} dx\, A_r \cdot A_m^* = \int_{-\infty}^{\infty} dy\{...\} \int_{-\infty}^{0} dx\, e^{-\frac{2x^2}{w_{eff}^2}} e^{ik\frac{x^2}{2R_{rel}}} e^{-ik\cdot x\cdot\varphi\left(1-\frac{z_{tm}}{R_m}\right)} e^{ik\cdot x\cdot\frac{x_{0s}}{R_m}} \quad (6)$$

where the effective beam $w_{eff}$, the relative beam radius $R_{rel}$, the "lever arm length" $z_{tm}$ and the static beam displacement $x_{0s}$ were introduced:

$$\begin{aligned}\frac{2}{w_{eff}^2} &= \frac{1}{w_m^2} + \frac{1}{w_r^2} \\ \frac{1}{R_{rel}} &= \frac{1}{R_r} - \frac{1}{R_m} \\ x_0 &= x_{0s} + \varphi \cdot z_{tm}\end{aligned} \quad (7)$$

Note that in Equation 6 we neglected the beam displacement $x_0$ of the measurement beam in the expression for the intensity distribution $\exp(-2x^2/w_{eff}^2)$. We have shown by comparison to the direct numerical integration of the exact expression that this is a very accurate approximation as long as the beam centers are not offset by more than half a beam waist. The overall distance between the two beams is the sum of a static distance ($x_{0s}$) when the two beams are parallel, and "the dynamically changing distance" ($\varphi \cdot z_{tm}$), the origin of which is the deflection of the measurement beam from its parallel path by the angle $\varphi$ at a distance $z_{tm}$ from the plane of interference.

In Equation 6 the terms in x and y separate and can therefore be integrated independently. As a consequence the integral in the vertical direction (y) cancels in the calculation of the DWS signal $DWS_\varphi$, which is defined as the differential phase between the left and the right half of the quadrant diode:

$$DWS_\varphi = Arg\left(\frac{F_{left}}{F_{right}}\right) = Arg\left(\frac{\int_{-\infty}^{\infty} dy \int_{-\infty}^{0} dx\, A_r \cdot A_m^*}{\int_{-\infty}^{\infty} dy \int_{0}^{\infty} dx\, A_r \cdot A_m^*}\right) =$$

$$Arg\left(\frac{\int_{-\infty}^{0} dx\, e^{-\frac{2x^2}{w_{eff}^2}} e^{ik\frac{x^2}{2R_{rel}}} e^{-ik\cdot x\cdot\varphi\left(1-\frac{z_{tm}}{R_m}\right)} e^{ik\cdot x\cdot\frac{x_{0s}}{R_m}}}{\int_{0}^{\infty} dx\, e^{-\frac{2x^2}{w_{eff}^2}} e^{ik\frac{x^2}{2R_{rel}}} e^{-ik\cdot x\cdot\varphi\left(1-\frac{z_{tm}}{R_m}\right)} e^{ik\cdot x\cdot\frac{x_{0s}}{R_m}}}\right) \quad (8)$$

Considering the similarities of the terms in numerator and denominator of Equation 8, the derivation of $DWS_\varphi$ is somewhat simplified and we obtain after some algebra:

$$DWS_\varphi = Arg\left(\frac{1+i\cdot\alpha}{1-i\cdot\alpha}\right)$$

$$\alpha = erfi\left(\frac{w_{eff}}{2^{3/2}}\left(k\varphi\left(1-\frac{z_{tm}}{R_m}\right) - k\frac{x_{0s}}{R_m}\right)\cdot\sqrt{\frac{1+i\sigma}{1+\sigma^2}}\right), \quad (9)$$

$$\sigma = \frac{kw_{eff}^2}{4R_{rel}}$$

where $erfi$ is the imaginary error function defined by $erfi(z) = -i\,erf(i\,z)$. We now assume that the angle $\varphi$ and the static displacement $x_{0s}$ are very small and therefore only retain terms to first order in those parameters. After expansion of the error-function to first order and some lengthy algebra we find:

$$DWS_\varphi = \sqrt{\frac{2}{\pi}} \left( k\, w_{eff}\, \varphi \left(1 - \frac{z_{tm}}{R_m}\right) - k\, \frac{w_{eff}\, x_{0s}}{R_m} \right) F(\sigma) + O(\varphi^2, x_{0s}^2)$$

$$F(\sigma) = \frac{1}{\sqrt{2}} \sqrt{\frac{1+\sqrt{1+\sigma^2}}{1+\sigma^2}}$$

(10)

$F(\sigma)$ in Equation 10 depends on the nonlinearity parameter $\sigma$, which in turn depends on the diameter and the relative wave-front curvature of the two interfering beams. When $\sigma$ goes towards zero, $F(\sigma)$ approaches 1. Typically, $\sigma$ is quite small (~2E-1), but can become significant under certain conditions. Therefore it should not be neglected. Expanding $F(\sigma)$ to second order in $\sigma$ we obtain $F(\sigma) \approx (1 - 3/8\sigma^2)$. We shall now replace the angle $\varphi$ between the two interfering beams by the angle $\phi$ of the rotatable mirror and consider that $2\phi = \varphi$. The coupling parameter $D_\phi^{m1}$ relates the differential tilt of mirror 1 ($\Delta\phi = \phi_1 - \phi_2$) to the differential change in observed DWS-signal, in other words it describes the linear slope with respect to $\phi$ of Equation 10:

$$D_\phi^{m1} = \frac{d(DWS_\phi)}{d\phi} = \sqrt{2\pi}\, \frac{4 w_{eff}}{\lambda} \left(1 - \frac{z_{tm}}{R_m}\right) F(\sigma) \qquad (11)$$

We see from Equation 11 that the coupling coefficient depends to first order on the effective beam waist $w_{eff}$, the measurement beam curvature $R_m$ and the lever-arm length $z_{tm}$. At higher order, when $F(\sigma)$ becomes large, it also depends on the relative beam curvature $R_{rel}$.

Figure 3 shows the impact successive approximations have made, going from the exact result of Equation 9 (solid line) to the approximate analytical result for small relative beam angles of Equation 10 (dotted lines), up to the linear approximation for small nonlinearity parameter $\sigma$. We used the parameter values $R_{rel} = 5.0 m, w_{eff} = 8.8 \times 10^{-4} m$ which are typical for a heterodyne interferometer such as the one in [8,9]. The approximations (dotted line) are given for values of $\phi = 200\,\mu rad, 350\,\mu rad$ and $600\,\mu rad$, corresponding to values of the "small-angle parameter" $\gamma = kw_{eff}\phi$ of $\gamma = 1.0, 1.8$, and 2.6, respectively. We also checked the validity of Equation 9 by comparison to the direct numerical integration of the interference integrals and found perfect agreement.

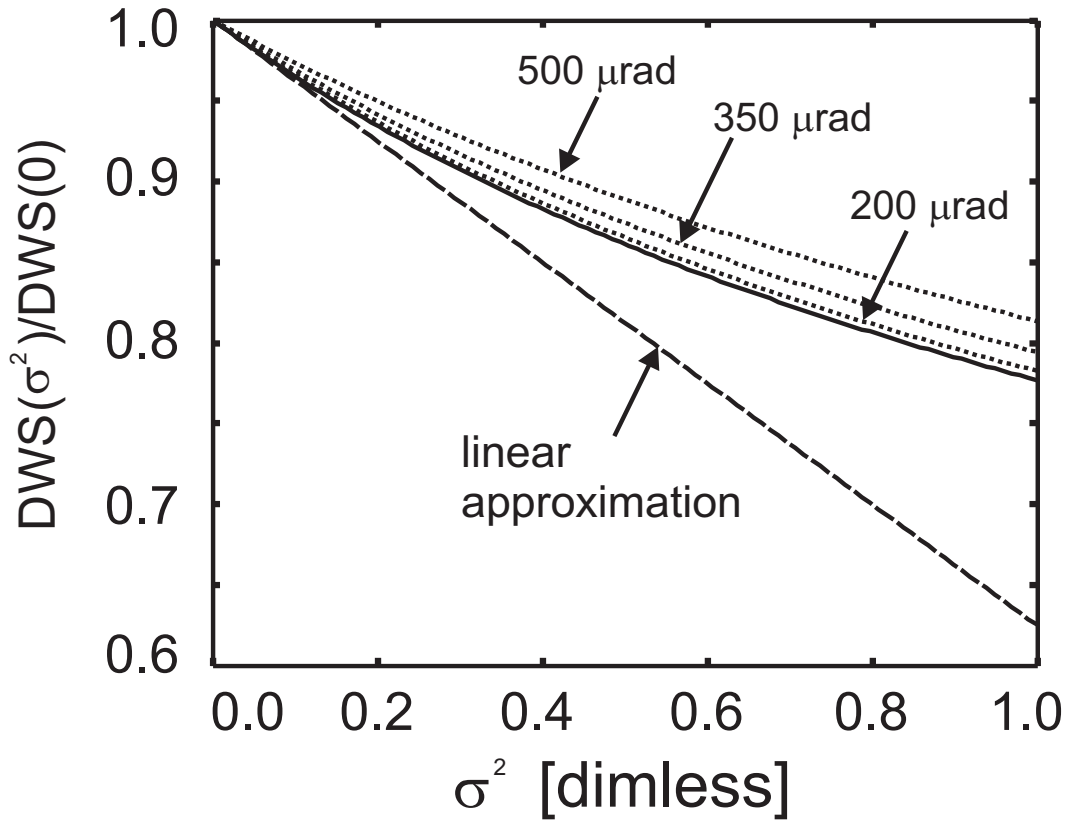

**Figure 3: The DWS-signals, normalized by their value for $\sigma^2 = 0$, are plotted against $\sigma^2$ while the value of $k\,w_{eff}\,\varphi$ is kept constant. The exact result is given by the solid line, the**

**approximate results are given by the dotted lines, and the linearized approximate result is given by the dashed line.**

We conclude that the approximation leading from Equation 9 to Equation 10 is a very good one for values of $\gamma \leq 1$. Equation 11 for the coupling parameter $D_\phi^{m1}$ provides the essential basis for the proposed measurements of the absolute beam curvature in the next section and it represents one major result of this article.

## Measuring the absolute wavefront curvature

### *Measurement approach*

The dependency of Equation 11 on the lever-arm length can be exploited by a setup such as the one depicted in Figure 1. We propose to perform a measurement where at first mirror 1 is rotated by the differential angle $\Delta\phi$ and the corresponding difference in DWS-signal is measured. In the next step mirror 2 is rotated by exactly the same differential angle $\Delta\phi$ and the corresponding difference in DWS-signal is measured. The two measurements completely determine the coupling parameters $D_\phi^{m1}, D_\phi^{m2}$. Upon inspection of Equation 11 we notice that all physical parameters determining the value of $D_\phi$ are identical in both test-cases except for the lever-arm length. Taking the ratio $\beta$ of the two coupling parameters we are therefore left with an expression that contains the radius of curvature of the measurement beam as the only unknown:

$$\beta = \frac{D_\phi^{m1}}{D_\phi^{m2}} = \frac{1 - z_{m1}/R_m}{1 - z_{m2}/R_m} \qquad (12)$$

Solving Equation 12 for the radius of curvature we find

$$R_m = \frac{z_{m1} - \beta z_{m2}}{1-\beta} \qquad (13)$$

We have therefore shown that through measurement of only two DWS-signal differences, when first mirror 1 and then mirror 2 are tilted, it is possible to determine the absolute wavefront curvature of the measurement beam.

*Error analysis*

The error in the ratio of coupling parameters ($\beta$) is given by

$$\Delta\beta = \frac{D_\phi^{m1}}{D_\phi^{m2}} \sqrt{\left(\frac{\Delta D_\phi^{m1}}{D_\phi^{m1}}\right)^2 + \left(\frac{\Delta D_\phi^{m2}}{D_\phi^{m2}}\right)^2}. \qquad (14)$$

Considering an error in the coupling coefficients of $\Delta D_\phi \approx 0.03 \times D_\phi$, which is typical for measurements where the mirror tilt is accomplished by piezo-electric transducers with inherently large inaccuracies and hysteretic behavior, we obtain $\Delta\beta \approx \beta \times 0.05$. Assuming typical values of $\beta = 2/3$, $R_m = 1.0\,m$ and $z_{m1} = 0.5\,m$ and $z_{m2} = 0.25\,m$, we find for the error $\Delta R_m$ of the radius of curvature:

$$\Delta R_m = \frac{\Delta\beta}{1-\beta}(R_m - z_{m2}) = 0.08\,m \qquad (15)$$

The finite wavefront curvature affects a change in phase across the beam which is given by the complex exponent of Equation 3. We find that the phase $\varphi(x)$ changes from the beam center to the edge of the waist by an amount

$$\varphi(w) = \frac{kw^2}{2R_m} \approx \frac{2\pi}{3} \triangleq \frac{\lambda}{3}, \qquad (16)$$

where we used a value of $w = 10^{-3} m$ for the beam waist and the wave-length $\lambda = 1.064 \times 10^{-6} m$. From the error in the measurement of $R_m$ (Equation 15) we find the error in the phase measurement:

$$\Delta\varphi(w) = \frac{kw^2}{2R_m} \frac{\Delta R_m}{R_m} \triangleq \frac{\lambda}{40} \qquad (17)$$

In other words, a measurement of the wavefront curvature with the accuracy given in Equation 15 implies a phase-measurement with the accuracy given in Equation 17.

Note that the overall accuracy depends on the precision with which the coupling parameters $D_\phi$ are measured. Those in turn depend on the precision with which the dummy mirrors 1 and 2 can be reproducibly rotated by the differential angle $(\phi_1 - \phi_2)$. If we assume perfect tilt precision, the measurement is only limited by the heterodyne phase-meter accuracy (typically better than $10^{-4} rad$) which implies a resolution limit of $\Delta R_m \approx 2 \times 10^{-4} m$.

## *Determining other beam parameters*

It is possible to determine all beam parameters in situ, i.e. from the setup described in Figure 1 and without recourse to additional test- or measurement equipment. This simplifies any efforts and is useful in instruments where either tight confinement or handling restrictions do not allow the controlled insertion of measurement probes on multiple locations of the beam path. On the downside, due to the strong dependence of $D_\phi$ on the measurement rather than the reference beam parameters, the values for the reference beam parameters cannot be determined very accurately by the method suggested below, but it suffices to verify that the calculated parameters are consistent with expectations.

In total we have 4 unknowns, the two beam parameters for the measurement beam $R_m, w_m$ and the two parameters for the reference beam, $R_r, w_r$, but only 2 Equations to solve for them. One of the unknowns, $R_m$, has already been calculated from the ratio of the two equations, which only leaves 1 independent equation. The waist of the measurement beam can be determined in a separate measurement by incrementally tilting mirror 1 and scanning the beam profile across the diode.

To find the beam parameters of the reference beam we revert to the full Equation 11 for $D_\phi^{m1}$ and the analogous equation for $D_\phi^{m2}$. Using the spare photo-diode located in the other measurement arm after the recombination beam splitter (dotted line in Figure 1), we obtain two more Equations for $\hat{D}_\phi^{m1}$ and $\hat{D}_\phi^{m2}$ which denote the coupling parameters for mirrors 1 and 2 measured on the spare diode, respectively. The beam waists and curvature radii are different on this diode to those on the principal diode but they can be related to another through Equation 2. The wave-front curvature radius of the measurement beam can be found once more from the ratio of $\hat{D}_\phi^{m1}$ to $\hat{D}_\phi^{m2}$. Using an iterative simplex search method, the two reference beam parameters are easily found from the two independent equations for $D_\phi^{m1}$ and $\hat{D}_\phi^{m1}$. From a simulation which implements the setup of Figure 1 and the measurements performed to determine the beam parameters, we find that the simulated measurement results match the input beam parameters with sufficient accuracy.

## Position dependence of the coupling parameters and noise coupling

Some parts of this section, especially the final paragraph, address specific sensitivity and alignment issues which are of importance in the design and usage of high-precision metrology systems such as those used in gravitational wave detectors [8,9,12,13].

In this section we investigate to which extent the coupling coefficients $D_\phi$ of Equation 11 change with increasing lever arm-length, e.g. through the movement of mirror 1 or 2 along the axis of beam propagation (z-axis). As a consequence, the beam curvature, beam waist and lever arm length change on the interference plane so that all parameters which enter Equations 10 and 12, namely $w_{eff}, z_m, R_m, R_{rel}$, are affected. In the following discussion we consistently use the same beam parameters as in the previous sections: $R_m = 1.0\,m, w_m = 1.0 \times 10^{-3}\,m$ and $R_r = 5/6\,m, w_r = 0.8 \times 10^{-3}\,m$. The lever-arm length is chosen to be $z_{m1} = 0.55\,m$.

For an interferometer such as the one in Figure 1 any test-mirror movement is accompanied by a parallel displacement of the measurement beam. A similar side-effect also occurs in the interferometers described in [8-11] where the beam is deflected from a "test-mirror" at an acute angle of $\alpha = 4.5\,\deg$. A simplified schematic is given in Figure 4, where we see that any change of the mirror position by the distance $\Delta z$ implies a beam displacement by the distance $d = \Delta z \sin \alpha$. After moving the mirror the beam can either be left as it is, that is displaced from its original position on the photo-diode (dashed line), or the mirror is tilted slightly by the angle $\varepsilon = d / z_{tm}$ so that the displaced beam is guided back (dotted line) onto the center of the photo-diode for optimal interference with the reference beam. For comparison the original beam path is given by the solid line.

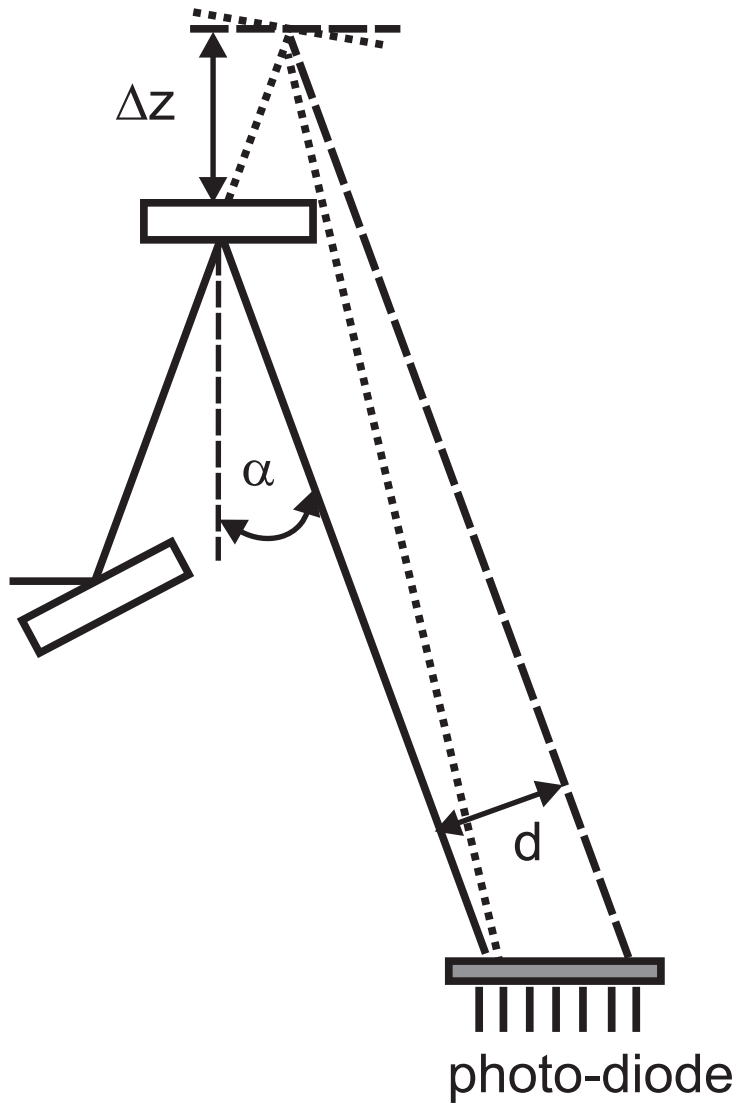

**Figure 4: Schematic of the interferometer described in [8]. The measurement beam is reflected from the test mirror at an angle $\alpha$ (solid line). When the mirror is moved the beam is displaced (dashed line). Its center on the photo-diode can be aligned with its original position when the mirror is slightly tilted (dotted line).**

The corresponding change in coupling coefficients, relative to the original value before the mirror was moved, is shown in Figure 5. The coefficient decreases rapidly with increasing beam displacement (dashed line). If the beam displacement is compensated by increasing the angle between measurement and reference beam (see Figure 4), the coefficient decreases similarly fast (dotted line). If a polarizing interferometer is used, as in the gravitational wave-detector LISA [12,13] which is currently being built, the measurement beam incidence is normal to the test-mirror and therefore no beam displacement occurs. There is only a very small increase of coupling constants due to the variation in beam width and wavefront curvature radius (solid line).

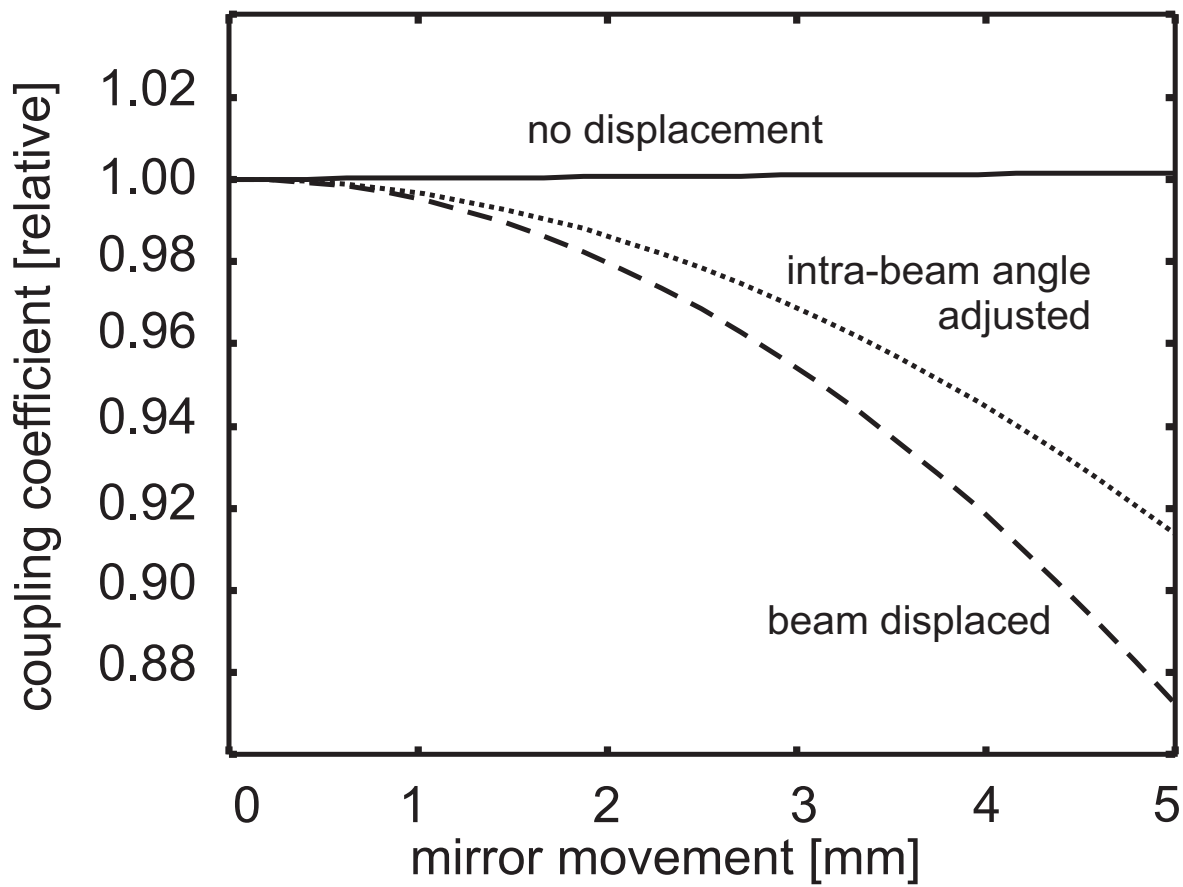

**Figure 5: The relative change in coupling coefficients $D_\phi$ when the test-mirror in the setup of Figure 4 is moved. The change in coefficients for increasing beam displacement is given by the dashed line. The dotted line is given for the case where the beam displacement is compensated by adjusting the test-mirror and increasing the angle between reference and measurement beam. The solid line is given for the case of $\alpha = 0$ (no beam displacement).**

We find for the slopes $s_\phi = D_\phi^{-1}(z=0) \cdot dD_\phi/dz$ of the solid and dashed lines the values of $+0.3\,m^{-1}$ and $-50\,m^{-1}$, respectively. The slope of the dashed line implies that the coupling coefficient changes by -5% per millimeter displacement, which stresses the importance of correct alignment of the test-mirror (in the actual detector it is a floating test-mass) with the nominal interferometer position.

The slope of the curves in Figure 5 also determines to what degree longitudinal jitter of the test-mass couples into measurement noise of the attitude. Real physical (as opposed to measurement sensor related) longitudinal displacement noise of linear spectral density $n_z$ couples into angular noise of linear spectral density $n_\phi$ as given by the following expression: $n_\phi = s_\phi \times \phi \times n_z$. We obtain $n_\phi = 2 \times 10^{-13}\,rad\,Hz^{-1/2}$ for the realistically chosen parameters $n_z = 10 \times 10^{-12}\,m\,Hz^{-1/2}$, $\phi = 400\,\mu rad$ and $s_\phi = 50\,m^{-1}$, which is 4 orders of magnitude below sensitivity requirements. In conclusion, even though an interferometer geometry where the measurement beam is reflected at an acute favors the coupling of longitudinal into angular noise, the expected effects are very small.

## Conclusion

We derived an exact analytical expression for the differential phase across the interference pattern of two Gaussian beams as detected by a quadrant diode. Starting from the exact expression we successively introduced further approximations to simplify the discussion and gain a detailed understanding of the parameter inter-dependencies. All approximations were justified in detail by numerical comparison to the exact expression.

Based on our findings we proposed a novel method to directly measure the absolute radius of wavefront curvature which requires two differential phase-measurements following the successive rotation of two test-mirrors. The measurement errors for this technique and its limiting accuracy were discussed in detail. We also studied the variation of the differential phase-measurements as a function of the test-mirror movement and calculated the change in coupling coefficient with increasing distance of the test-mirror from its nominal location. A consequence of this effect, the coupling of longitudinal displacement noise into differential-phase noise, was shown to be negligible in typical heterodyne interferometers.

## Acknowledgements

G.H. gratefully acknowledges stimulating discussions with David Robertson, University of Glasgow, Gudrun Wanner, Albert Einstein Institut in Hannover, Vinzenz Wand, Rüdiger Gerndt and Ulrich Johann, EADS Astrium Friedrichshafen. G.H. is grateful to I.C.W. from the R.B.C for providing the encouragement, support and means to write this paper.

.